\documentstyle[aps,aipbook,floats,epsf]{revtex}
\begin{document}

%  Aliases

\def\beq{\begin{equation}}
\def\eeq{\end{equation}}
\def\bea{\begin{eqnarray}}
\def\eea{\end{eqnarray}}
\def\Csix{C^{(6)}}
\def\Ceight{C^{(8)}}
\def\costhstar{\cos\theta^*}
\def\Osix{O^{(6)}}
\def\Oeight{O^{(8)}}
\def\tbar{\bar t}
\def\shat{\hat s}
\def\that{\hat t}
\def\uhat{\hat u}
\def\CA{{\cal A}}
\def\GeV{{\>\, \rm GeV}}
\def\mtsq{m_t^2}
\def\third{{1\over3}}
\def\half{{1\over2}}
\def\mt{m_t}
\def\mperp{{m_\perp}}
\def\qbar{\bar q}
\def\pperp{p_\perp}
\def\Rang{R_{\rm ang}}
\def\Rrms{R_{\rm rms}}
\def\TeV{{\>\, \rm TeV}}

\title{Anomalous Gluon Self-Interactions\\
and $t\bar t$ Production}

\author{Elizabeth H. Simmons$^*$ and Peter Cho$^{**}$}
\address{$^*$Department of Physics, Boston University \\
590 Commonwealth Avenue, Boston MA 02215 \\
$^{**}$ Lauritsen Laboratory, California Institute of Technology \\
Pasadena, CA 91125 }

\maketitle

\begin{abstract}
Strong-interaction physics that lies beyond the standard model may
conveniently be described by an effective Lagrangian.  The only
genuinely gluonic CP-conserving term at dimension six is the
three-gluon-field-strength operator $G^3$.  This operator, which
alters the 3-gluon and 4-gluon vertices form their standard model
forms, turns out to be difficult to detect in final states containing
light jets.  Its effects on top quark pair production hold the
greatest promise of visibility.
\end{abstract}

\section*{Introduction}

The hallmark of a non-abelian gauge theory is the self-interaction of
its gauge fields.  Most of this conference has been devoted to the
electroweak vector bosons; this talk\footnote{Presented by
E.H. Simmons.} will focus, instead, on the self-interactions of the
gluons.  Any experimental indication that the gluon self-coupling
differed from the form predicted by the $SU(3)$ gauge theory of the
strong interactions would point to the existence of new color-related
physics.  This talk presents a model-independent effective Lagrangian
analysis of possible non-standard contributions to color physics, and
assesses the possibility of measuring the coefficients of the
effective Lagrangian.

Suppose that some exotic color physics exists at an energy scale
$\Lambda$.  For instance, there might be new colored scalars or
fermions with a mass of order $\Lambda$, such as squarks and gluinos
\cite{ssusy}, colored technihadrons \cite{techni}, or fermions in
non-fundamental representations of $SU(3)$.
Or instead, perhaps the gluons and quarks are manifestly composite
\cite{compness} when probed at a distance scale $\Lambda^{-1}$.  Such
non-standard physics would lead to new gluon self-interactions through
virtual loops of heavy colored particles or through exchange of
sub-components.

A {\it complete} description of a given set of new phenomena would
require a fundamental theory beyond the standard model. But at low
energies $E << \Lambda$, where the underlying preon exchange or loops
of new particles cannot be resolved, the new color physics causes
multi-gluon contact interactions suppressed by inverse powers of
$\Lambda$.  These contact interactions are described by an effective
Lagrangian
\beq
{\cal L}_{eff} = {\cal L}_{QCD} + {1 \over \Lambda^2} \sum_i \Csix_i(\mu)
\Osix_i(\mu) + {1\over \Lambda^4} \sum_i \Ceight_i(\mu) \Oeight_i(\mu) +
O\bigl({1 \over \Lambda^6} \bigr)
\eeq
that includes the conventional QCD Lagrangian plus non-renormalizable
operators $O_i$ that are constructed from gluon field strengths
$G^{\mu\nu}$ or color-covariant derivatives $D^\mu = \partial^\mu - i
g_s G^\mu$.  The new operators obey the gauge and global symmetries
of the standard model.

Our task is to identify the leading operators in this effective
Lagrangian and determine which experiments are best able to detect
their effects.

\section*{leading operators in L$_{eff}$}

Since a non-renormalizable operator of dimension $(4+d)$ is suppressed
by $\Lambda^{-d}$, the operators making the most visible contribution
to physical processes will be those of lowest dimension.

The number of nonrenormalizable terms which arise at dimension 6 in
the gluon sector is small.  One can build only two gauge-invariant
operators preserving $C$, $P$ and $T$ out of covariant derivatives and
gluon field strengths \cite{ehsone}:
\bea
\Osix_1 &=& g_s f_{abc} G^\mu_{a\nu} G^\nu_{b\lambda} G^\lambda_{c\mu}
\label{dimsixopsa} \\
\Osix_2 &=& {1 \over 2} D^\mu G^a_{\mu\nu} D_\lambda G^{\lambda\nu}_a.
\label{dimsixopsb}
\eea
The triple gluon field strength
term in Eq. \ref{dimsixopsa}, which we shall name $G^3$ for short,
represents a true gluonic operator, contributing to three-gluon and four-gluon
non-abelian vertices.  The double gluon field strength operator in Eq.
\ref{dimsixopsb}, which we will call $(DG)^2$, is not really gluonic
in this sense.  The classical equation of motion
\beq
D_\mu G^{\mu\nu}_a = -g_s \sum_{\rm flavors} \bar q \gamma^\nu T_a q
\eeq
relates its S-matrix elements to those of a color octet four-quark operator
\cite{Politzer}:
\beq
\Osix_2
\> {\buildrel {\scriptscriptstyle EOM} \over \longrightarrow} \>
{g_s^2\over 2} \sum_{\rm flavors}
\bigl( \bar q \gamma_\mu T_a q  \bigr) \bigl( \bar q \gamma^\mu T_a q \bigr).
\eeq
The two-field-strength operator thus affects parton processes involving
external quarks rather than external gluons.

The list of $CP$-even gluon operators grows significantly at
dimension eight.  Classifying the operators according to the number of
field strengths that they contain, we find one independent operator
built from two field strengths and four covariant derivatives, two
operators with three gluon field strengths and two derivatives, and a
half-dozen operators containing four field
strengths\cite{ehstwo,running}.  Rather than listing all nine
operators explicitly (for a list, see \cite{ehstwo}) we merely mention
that there are two situations in which dimension-8 operators may give
noticeable effects.  One of the two-field-strength
operators,
\beq
\Oeight_3 = g_s f_{abc} G^\mu_{a\nu} G^\nu_{b\lambda} D^2 G^\lambda_{c\mu}.
\label{dimeightops}
\eeq
contributes at tree-level and order $1/\Lambda^4$ to the process $gg
\to q\bar q$; it is the only $d=8$ gluonic operator to do so.  The
effect of this operator on angular distributions will feature in our
discussion of $gg \to t\bar t$.  In addition, the four-field-strength
operators contribute significantly to the gluon four-point vertex and,
hence, the process $gg \to gg$ which we will analyze shortly.

\section*{Dijet production}
Having established our operator basis, we consider how best
to detect the presence of non-standard gluon self-interactions.  In
principle, the operator $G^3$, being the lowest-dimension operator to
affect multi-point gluon vertices, is the one to focus on.  A logical
beginning is to consider its effects on hadronic scattering at
high-energy colliders like FNAL and the LHC.  Because scattering at
both colliders is dominated by gluon-gluon collisions, one expects
non-standard gluon self-interactions to noticeably affect two-body
scattering cross-sections.  These cross-sections, in turn, dominate
the well-measured (at FNAL) inclusive single-jet production
cross-section.  Hence, it appears that a strong limit on the coefficient
$\Csix_1 / \Lambda^2$ should be forthcoming.

The leading contribution of the $G^3$
operator to the inclusive jet cross-section ${\bf p\, p\!\!\!\!\!^{(-)} \to
jet + X}$ is expected to lie in its effect on the sub-process
$gg \to gg$.  Consider the Feynman diagrams contributing to the $gg
\to gg$ scattering amplitude in pure QCD and those with one insertion
of $G^3$ (i.e., one anomalous multi-gluon vertex per diagram).  The
ordinary QCD contribution to the scattering comes from squaring the
sum of the QCD diagrams; the lowest-order ($1\over \Lambda^2$) piece
due to $G^3$ arises from the interference of the QCD and one-insertion
diagrams.  It has been shown \cite{ehsone}, however, that the QCD
amplitude is only in the $[++++]$ helicity channel and the
one-insertion amplitude is purely in the orthogonal $[++--]$ and
$[+---]$ channels.  There is, consequently, no order $1\over\Lambda^2$
contribution to $gg \to gg$.  The leading effect of the $G^3$ operator
arises at order $1/\Lambda^4$, from squaring the one-insertion
diagrams and from interfering two-insertion diagrams with the QCD
diagrams.

Where a lower-order effect is missing, one would hope to
experimentally detect the remaining higher-order effect.  This will be
difficult. The leading contributions of the dimension-eight
four-field-strength operators to $gg \to gg$ also arise at order
$1/\Lambda^4$ when an amplitude with one insertion of a
dimension-eight operator interferes with a QCD amplitude.
Furthermore, the order $1/\Lambda^4$ contributions to gluon scattering
of the operators $G^3$ and $f_{eab} f_{ecd} G^{\mu\nu}_a
G^{\lambda\rho}_b G_{c\mu\nu} G_{d\lambda\rho}$ are identical in form
and similar in magnitude\cite{ehstwo}.  Isolating the effects of $G^3$
in $gg\to gg$ does not appear possible.

The next most promising sub-processes appear to be
those involving massless quarks as well as gluons, i.e.
$g q \to g q$ and reactions related to it by crossing. Initial state
gluons are still a possibility and $G^3$ can enter some
diagrams through the three-gluon vertex.  When the contribution of
$G^3$ is calculated, however, the order $1/\Lambda^2$ piece vanishes.
The leading effects are,again, order $1/\Lambda^4$ and compete
with the effects of higher-dimension operators.

In summary: two-body scattering of massless partons does not put
significant limits on non-standard gluon self-interactions involving
the $G^3$ operator.  Because the leading effects are of order
$1/\Lambda^4$, there is competition from higher-dimension operators.
Furthermore, the fact that gluons predominate at low $x$ where the QCD
background is greatest weakens the attainable bounds\cite{ehstwo}.

\section*{Other jet-production experiments}
While the $G^3$ operator cannot be detected in $2 \to 2$ light parton
scattering processes, there are several other options for studying
non-standard strong interactions using massless final-state jets.  As
will become clear, none is fully satisfactory.

Although $G^3$ does not affect dijet production at tree level, the other
dimension-six operator, $(DG)^2$ can. As noted earlier, this operator
alters the propagator and coupling of internal gluons.  At leading
order its effects on scattering are equivalent to those of the
four-quark operator $\left({\bf \bar\psi \gamma^\mu} T^a
{\bf \psi}\right)^2$. Consider, for example, the process ${\bf p\bar p \to
jet + X}$, to which dijet production is the leading contributor.  At
FNAL, initial quarks play a larger role than initial gluons in
scattering at large Bjorken $x$; thus high $\pperp$ jets are more likely
to originate from initial quarks.  The inclusive jet cross-section is
found to fall more slowly with $\hat s$ or $\pperp$ when an insertion of
the $(DG)^2$ operator is included than when only QCD is studied
\cite{ehsone,ehstwo,ehspcone}.  This makes the transverse-momentum
spectrum potentially sensitive to the presence of the $(DG)^2$
operator.  Analyzing published CDF inclusive jet data \cite{fnalfq}
yields a lower bound of 2 TeV on the scale of new physics associated
with this operator\cite{ehspcone}.  This limit is useful -- but
because no external gluons are involved, it is only tangentially
related to probing the structure of the gluon self-coupling.

At LEP, one source of 4-jet final states is the process wherein a $Z$
decay to a quark/anti-quark pair is followed by radiation of a gluon
that splits into a pair of gluons.  The three-gluon vertex involved in
this process can be affected by the presence of the $G^3$ operator.
In \cite{ddz,dufzep} it was found that while most kinematic
variables describing $Z \to 4 j$ would reflect the presence of $G^3$ only in
the overall {\it rate}, the dijet invariant mass distribution of the two
most energetic jets changes {\it shape} when the $G^3$ operator is
included.  With 10pb$^{-1}$ of data, it should be possible to set a limit of
$\Lambda >$ 100 GeV using this dijet invariant mass distribution; ten
times the data would boost the limit to 175 GeV.  The limiting factor
is the energy at which LEP experiments are performed.

While the order $1/\Lambda^2$ contributions of the $G^3$ operator to dijet
production vanish at tree level, the same is not true when larger numbers
of jets are being produced.  The very difference in helicity properties
between the scattering amplitudes of pure QCD and those with one insertion
of $G^3$ which keeps the $2 \to 2$ amplitudes from interfering provides a
potential signal of the presence of $G^3$ in $2 \to 3$ processes
\cite{dixon}.  For example, if one considers $g g \to g g g$ when two of
the outgoing gluons are nearly collinear one observes the following.
Treating the nearby gluons as one effective gluon yields an approximate $2
\to 2$ process for which we know the helicity properties of scattering
amplitudes with and without $G^3$.  The pure QCD amplitude with its $[++++]$
helicity is symmetric under azimuthal rotations about the momentum vector
of the effective gluon. The amplitude with an insertion of $G^3$ admits the
$[++--]$ and $[+++-]$ helicities which allows the effective gluon to be
linearly polarized, yielding azimuthal dependence.  No limits on $\Lambda$
have yet been suggested using this method; the limiting factor may be the
experimental difficulty of studying 3-jet events in the near-collinear
region.

\section*{The top quark production cross-section}

Light jets having failed us, we turn to the possibility of studying
anomalous gluon self-interactions through their effects on scattering
involving heavy flavors.  The two-body scattering process that both
involves heavy fermions and benefits maximally from the high gluon
luminosity at hadron colliders is $g g \to q \bar q$.  We will find
that the order $1/\Lambda^2$ contribution of $G^3$ to this process is
proportional to $m_q^2$; hence the effect is greatest for top
production.  The top quark is also the easiest to tag since its
leptonic decay channel can produce a high energy, isolated lepton in
conjunction with a bottom quark.  This distinctive signature cuts down
on genuine backgrounds as well as false identifications.  To the
extent that $b\bar{b}$ and even $c\bar{c}$ final states can be cleanly
identified, the signal for $G^3$ will be enhanced and our results can
be applied to their study.

To study top quark production, we must enlarge our basis of
higher-dimension operators.  The gluonic operators described above are
not the only higher-dimension operators that can affect top
production.  They also do not form a closed basis under one-loop
renormalization, which we employ in running down from the
scale of new physics to the top production threshold.
The operators $\Osix_1$ and $\Osix_2$ do not mix with each other under
the action of QCD at one-loop order.  Instead, $\Osix_1$ runs into
itself and the chromomagnetic moment operator
\beq
\Osix_0 = \sum_{\rm flavors} g_s m_q
  \bar q\sigma^{\mu\nu} T^a q G^a_{\mu\nu}.
\eeq
Because the equations of motion relate $\Osix_2$ to a color-octet
four-quark operator, $\Osix_2$ mixes at one loop with other
four-quark operators
\bea
\Osix_3 &=& {g_s^2 \over 2} \sum_{\rm flavors}
   \bigl( \bar q\gamma_\mu \gamma^5 T_a  q \bigr)
   \bigl( \bar q\gamma^\mu \gamma^5 T_a q \bigr) \\
\Osix_4 &=& {g_s^2\over 2} \sum_{\rm flavors}
   \bigl( \bar q\gamma_\mu q \bigr) \bigl( \bar q\gamma^\mu q \bigr) \\
\Osix_5 &=& {g_s^2\over 2} \sum_{\rm flavors}
   \bigl( \bar q\gamma_\mu \gamma^5 q \bigr) \bigl( \bar q\gamma^\mu
\gamma^5 q \bigr)
\eea
The mixing matrices are given explicitly in \cite{ehspctwo,running}.

Renormalization group evolution suppresses the
coefficients $\Csix_1$ and $\Csix_2$.  Given that one expects some new
fundamental layer of physics to lie in the TeV regime, we will take
$\Lambda = $2 TeV.  This ensures that our analysis will be valid over
almost the entire energy range of present and anticipated
hadron colliders.  If the operator coefficients assume the values
\beq
(\Csix_0,\Csix_1,\Csix_2,\Csix_3,\Csix_4,\Csix_5)(\Lambda)
= (1,1,1,0,0,0)
\eeq
at a scale of 2 TeV, then they run down to
the values
\beq
(0.7858, 0.7458, 0.8856, -0.0294, 0.0003, -0.0152)
\eeq
at the top-antitop threshold\cite{ehspctwo}.

The partonic sub-processes contributing to top quark pair-production
at a hadronic collider are $gg \to t\bar t$ and $q \bar q \to t\bar
t$.  Both the $G^3$ and chromomagnetic moment operators contribute to
the gluon fusion channel; the non-standard contribution to the
squared matrix element is
\bea
\bar{{\sum}^\prime} &\vert& {\cal A}(gg \to t\tbar) \vert^2 =
 {\mtsq \over \Lambda^2} \Bigl[ {{\Csix_0
   ({4\over3} \shat^2-3 \that\uhat-3\mtsq\shat+3 \mt^4) + {9\over8}
\Csix_1  (\that-\uhat)^2} \over
   {(\mtsq-\that) (\mtsq-\uhat)}}\Bigr] \nonumber \\
&+& {1 \over \Lambda^4} \Bigl[
   {1 \over 6} {\Csix_0}^2 {\mtsq \bigl( 14 \shat\that\uhat
   + \mtsq(31 \shat^2-36 \that\uhat) - 50 \mt^4 \shat + 36 \mt^6 \bigr) \over
   (\mtsq-\that)(\mtsq-\uhat)} \label{gluesqrdamp}\\
&+&{9 \over 8} \Csix_0 \Csix_1 {\mtsq \shat^3 \over
   (\mtsq-\that)(\mtsq-\uhat)}
+ {27 \over 4} {\Csix_1}^2 (\mtsq-\that)(\mtsq-\uhat) \Bigr]
+ O\Bigl({1 \over \Lambda^6} \Bigr).\nonumber
\eea
where the bar over the $\Sigma$ implies averaging
(summing) over initial (final) spins and colors,
while the prime indicates division by $g_s^4$.
Notice that all of the nonrenormalizable operator
terms except the last one are proportional to
$\mtsq$.  Because the last term is enhanced by a prefactor of
$27/4$ and increases quadratically with $\shat$, well
away from the $t\tbar$ threshold and over large
regions of $\Csix_1$ parameter space, the $\Osix_1$
operator's squared amplitude is much larger than
its interference with QCD.

	One may question whether the $O(1/\Lambda^4)$
terms arising from dimension-8 gluon operators could be significant.
The answer is generally no.  The only dimension-8
operator that
affects $gg \to t\bar t$ scattering at lowest order is $\Oeight_3$
\beq
\bar{{\sum}^\prime} | \CA(gg \to t\tbar) |^2 = \cdots
- - -{3 \over 8} {\Ceight_3 \over \Lambda^4} {\mtsq \shat(\that-\uhat)^2
\over(\mtsq-\that)(\mtsq-\uhat)}.
\label{Oeightsqrdamp}
\eeq
This term has a smaller prefactor and increases more slowly with
$\shat$ than the term proportional to $(\Csix_1)^2$ and is unlikely to
obscure any signal from $\Osix_1$.

The $(DG)^2$, chromomagnetic moment, and four-quark operators make
the following addition to the quark/anti-quark annihilation  matrix element
\bea
\bar{{\sum}^\prime} &|&\CA(q \qbar \to t\tbar) |^2 =  \label{quarksqrdamp}\\
 &&{1\over 9 \shat \Lambda^2}\bigl[4 \Csix_0 \mtsq \shat +
\Csix_2 (\that^2+\uhat^2+4\mtsq\shat-2\mt^4)
  + \Csix_3 \shat (\that-\uhat) \bigr] \nonumber \\
&+& { 4 \over 9 \Lambda^4} \Bigl[ 8 {\Csix_0}^2 \mtsq
(\that\uhat+2\mtsq\shat-\mt^4)/\shat
  + 8 \Csix_0 \Csix_3 \mtsq(\that-\uhat)  \nonumber \\
&+& 8 \Csix_0 \Csix_2 \mtsq \shat
+ ({\Csix_2}^2+ \half {\Csix_4}^2)
  (\that^2+\uhat^2+4\mtsq\shat-2 \mt^4) \nonumber \\
&+& ({\Csix_3}^2+\half {\Csix_5}^2)
  (\that^2+\uhat^2-2 \mt^4 )
+ (2\Csix_2 \Csix_3 + \Csix_4 \Csix_5) \shat(\that-\uhat) \nonumber
\Bigr].
\eea
with the same conventions as before.  Unlike Eq. \ref{gluesqrdamp}
this expression contains no anomalously large order
$1/\Lambda^4$ term.  So we expect the effect of dimension-eight
and higher operators upon $q\bar q \to t\bar t$ scattering to be small
\cite{ehspctwo}.

	The squared amplitudes in Eqs. \ref{gluesqrdamp} and
\ref{quarksqrdamp} enter the partonic cross section
\beq
{d \sigma(ab \to t\tbar) \over d\that} = {\pi \alpha_s^2 \over \shat^2}
\bar{{\sum}^\prime} | \CA(ab \to t \tbar) |^2 \ \ .
\label{ggttbarXsect}
\eeq
This is combined with distribution functions
$f_{a/A} (x_a)$ and $f_{b/B}(x_b)$ specifying the probability of finding
partons $a$ and $b$ inside hadrons $A$ and $B$ carrying momentum fractions
$x_a$ and $x_b$ and summed over initial parton
configurations.  The resulting hadronic cross section
\beq
{d^3\sigma \over dy_3 dy_4 d\pperp} \bigl(AB \to t \tbar \bigr)
= 2 p_\perp \sum_{ab} x_a x_b f_{a/A}(x_a) f_{b/B}(x_b)
{d\sigma (ab\to t\tbar)\over d\that}.
\label{ABttbarXsect}
\eeq
depends on the top and antitop rapidities $y_3$ and $y_4$ and their
common transverse momentum $p_\perp$.

\section*{Top quark production and gluon self-interactions}

We now compare the effects of various non-standard gluon interactions
on the top quark production cross-section.  This discussion will
communicate the conclusions that can be drawn from the salient
features of the kinematic distributions of the top quarks.  Full
details reside in \cite{ehspctwo}.

We first examine the transverse
momentum distribution obtained by integrating
$d^3 \sigma / dy_3 dy_4 d\pperp$ over the rapidity range
$-2.5 \le y_3, y_4 \le 2.5$.
\footnote{This convenient integration
interval contains the bulk of the produced
top quarks.  Extending the range to
$-6 \le y_3, y_4 \le 6$ does not alter our results.}
The resulting $\pperp$ distribution of $t\tbar$ pairs produced at the LHC
is plotted in Fig. 1, which shows curves for QCD and for the separate
contributions of operators $\Osix_0$, $\Osix_1$ and $\Osix_2$ with
their respective $C_i(\Lambda)$ equal to 0.5.
\footnote{We used the next-to-leading order parton
distribution function set B of Harriman, Martin, Roberts and Stirling
\cite{Harriman} evaluated at the
$\mu=\mperp\equiv\sqrt{\mtsq+\pperp^2}$.}
%
%
%%%
%%%FIG-A GOES HERE
%%%
\begin{figure}[ht] % fig 1
%\vspace*{1in}
\epsfxsize=4.7in
\epsfbox[0 0 504 340]{figa.ps}
\caption[figone]{ $d\sigma(pp \to t\tbar)/d p_\perp $ at the LHC with
$\sqrt{s} = 14 \TeV $.  The solid curve represents pure QCD.  The
dot-dashed, dashed and dotted curves show the additional contributions
when either $\Csix_0(\Lambda)$, $\Csix_1(\Lambda)$ or
$\Csix_2(\Lambda)$ is set to 0.5 with $\Lambda=2\TeV $.}
\end{figure}

The $\pperp$ dependence of the curves in Fig. 1
differentiates the dimension-6 operators
from each other and from QCD terms in ${\cal
L}_{eff}$. At high $\pperp$, where the
QCD background is lowest, the dimension-6
operator making the largest contribution to the
rate of top quark production is $G^3$.  Next in
importance at large $\pperp$ is $(DG)^2$; the
chromomagnetic magnetic operator lags far behind.
Placing a lower $\pperp$ cut around 500 GeV,
can eliminate most of the chromomagnetic moment
operator's contribution in favor of that from $G^3$
and $(DG)^2$.  In addition, the shapes of the
curves for $G^3$ and $(DG)^2$ are noticeably
different from one another and from the shape of
the QCD curve; that of the chromomagnetic moment
operator closely mimics QCD.  Hence the $G^3$
operator should make the most visible contribution
to $d\sigma/d\pperp$ at the LHC.

A quantitative comparison of the QCD and effective lagrangian
predictions for $d\sigma (pp \to t\tbar)/d\pperp$ at the LHC confirms
this \cite{ehspctwo}.  To compare rates, we computed the ratio
$R_\perp$ of the integrals of $d\sigma_{EFT}/d\pperp$ and
$d\sigma_{QCD}/d\pperp$ over the momentum range $500 \GeV < \pperp <
1000 \GeV$.  We found that $R_\perp$ was fairly insensitive to
$\Csix_0$, and depended a few times more strongly on $\Csix_1$ than on
$\Csix_2$.  To compare shapes, we formed a $\chi^2$
function for the difference in number of high-$\pperp$ events
predicted by QCD and ${\cal L}_{eff}$.  Again, the strongest
dependence was on $\Csix_0$.  Further, $R_\perp$ and the $\chi^2$
function depend differently on $\Csix_1$ and $\Csix_2$, making the
combined measurements even more powerful.  We estimate
that the LHC could set a limit $\vert \Csix_1 \vert < 0.5$.

The Tevatron analogues of the LHC differential
cross-sections are shown in Fig. 2.
%%%
%%%FIG-B GOES HERE
%%%
\begin{figure}[ht] % fig 2
%\vspace*{1in}
\epsfxsize=4.7in
\epsfbox[0 0 504 330]{figb.ps}
\caption[figtwo]{$d\sigma(p{\bar p} \to t\tbar)/d\pperp$ at FNAL with
$\sqrt{s} = 1.8 \TeV$.  The curves are labeled as in Fig. 1.}
\end{figure}
The integrated cross-section
for $t\tbar$ production is two orders of magnitude lower at the
Tevatron.  And the relative importance of the dimension-6 terms in the
effective Lagrangian depends upon $\sqrt{s}$ in a manner consistent
with the parton content of the colliding hadrons.
The most important dimension-six operators at
Tevatron energies are the chromomagnetic moment
operator $\Osix_0$ and four-quark operators like
$(DG)^2$; the effects of $G^3$ are (for equal
values of the $C_i$ at high energies) an order of
magnitude smaller.   Hence FNAL experiments are
unlikely to find evidence for the $G^3$ operator in
$t\bar t$ production.  The effects of $\Osix_0$ and
$\Osix_2$ may be visible in terms of enhanced
production {\it rate}; again, the fact that the shape of
the $\Osix_0$ curve is identical to that of the QCD
curve will make $\Osix_0$ more difficult to detect.

We next study the angular distribution of the
produced top quarks, $d\sigma(pp \to t\tbar)/d
\costhstar$, where $\theta^*$ denotes the angle
between the direction of the boost and that of the
top quark in the parton center-of-mass frame.
To enhance the signal we have imposed the cut $\pperp \ge 500 \GeV$.  We
also required the lab frame angle between the $t$ or $\tbar$ and the
beamline to exceed $25.4^\circ$; this ensures that the
pseudorapidities of the decay products from high momentum tops will
predominantly fall within $-2.5 \le \eta \le 2.5$, the approximate
acceptance of planned LHC detectors .

The angular distribution
is plotted in Fig. 3 for pure QCD and for QCD {\it plus} some of the
$O_i$.
%%%
%%%FIG-C GOES HERE
%%%
\begin{figure}[ht] % fig 3
\vspace*{1.27in}
\epsfxsize=4.5in
\epsfbox[36 150 576 440 ]{figc.ps}
\caption[figthre]{$d\sigma(pp \to t\bar t)/d\cos\theta^*$ at the LHC with
$\sqrt{s} = 14$ TeV.  The solid curve shows pure QCD.  The dotted curve
shows QCD plus $\Osix_1$ with
$\Csix_1(2 \TeV)=0.5$.  The dashed (dot-dashed)
curve shows QCD plus $\Oeight_3$ with
$\Ceight_3(2 \TeV)$ = 0.5 (-0.5).}
\end{figure}
The curves indicating the effects of the chromomagnetic moment
operator are not included in the figure because they closely trace the
QCD curve for $\Csix_0(\Lambda) = \pm 0.5$.  Likewise, the curves for
$\Csix_1(\Lambda) = -0.5$ and $\Csix_2(\Lambda) = 0.5$ are nearly
indistinguishable from the curve shown for $\Csix_1(\Lambda) = 0.5$.
The dimension-8 gluon operator $\Oeight_3$ induces deviations
\footnote{The coefficient $\Ceight_3$ was not evolved using the
renormalization group but was instead simply fixed at its $\Lambda$
scale value.}
from pure QCD which are clearly visible
in $d\sigma/d\costhstar$.  This is quite interesting since the effect
of $\Oeight_3$ upon the $t\tbar$ transverse momentum distribution was
negligible.  Indeed, we omitted the effects of $\Oeight_3$ from Fig. 1
and Fig. 2 since they would have been suppressed relative to the
dimension-6 operators' effects by more than an order of magnitude.

As in the analysis of the $\pperp$ distributions, we distinguish
between effects on the rate and the shape of the curves.
In discussing rate, we compare the integral of a given curve (with
respect to $\costhstar$) to that of the QCD curve, denoting the ratio
by $\Rang$.  A curve's shape is compared with that of the QCD curve by
forming the ratio ($\Rrms$) of the respective root-mean-squared values
of $\costhstar$.  For the curves arising when $\Csix_0(\Lambda)$,
$\Csix_1(\Lambda)$, $\Csix_2(\Lambda)$ or $\Ceight_3(\Lambda)$ is set
equal to 0.5, we find $\Rang = (1.03, 1.23, 1.46, 0.82)$ and $\Rrms =
(0.999, 0.978, 0.991, 0.871)$.  For analogous curves with the
$\Lambda$ scale coefficients set equal to -0.5, we find $\Rang =
(0.969, 1.23, 0.735, 1.18)$ and $\Rrms = (1.00, 0.978, 1.03, 1.08)$.
The most striking implication is that the dimension-8 gluon operator
alters the shape of the $t\tbar$ angular distribution more than
any dimension-6 operator in ${\cal L}_{eff}$ for comparable
values of the $C_i$.  The magnetic moment operator's angular
distribution is indistinguishable from that of pure QCD, while the
distributions of the $G^3$ and $(DG)^2$ operators differ significantly
from that of QCD in $\Rang$ but not in $\Rrms$.

Figure 4 summarizes the detectability of the operators we
have studied.  Each operator produces visible effects in a unique
combination of experiments.
%%%
%%%TAB-1 GOES HERE
%%%
\begin{figure}[ht] % fig 4
\vspace*{1.in}
\epsfxsize=4.7in
\epsfbox[70 480 550 600]{tab1.ps}
\caption[figfour]{Experiments able to detect each type of non-standard
gluon interaction.}
\end{figure}
\vspace*{-0.1in}

\section*{conclusions}

Anomalous gluon self-interactions are elusive.  Detecting them in dijet
production is nearly impossible.  Other measurements involving light jets
are energy-limited or intrinsically difficult.

Heavy flavor production may offer the best hope of seeing non-standard
gluon self-couplings.  While only $t\bar t$ production is analyzed
here, $b\bar b$ production should show similar effects.  A strong signal will
be provided by the shape of the transverse-momentum distribution of the
produced heavy quarks; non-standard strong interactions can visibly affect
the number of events at high transverse momentum.  The angular
distribution of the heavy fermions can also help discriminate among the effects
of different higher-dimension operators.

Top-quark pair production at the LHC will test the three-gluon vertex
well.  The contribution of the $G^3$ operator to the transverse-momentum
spectrum exceeds that of all other contact operators for similar values of
their coefficients.  For a scale of new physics $\Lambda = 2$TeV, LHC
experiments should be able to set an upper bound of $0.5$ on the
coefficient of the $G^3$ operator.  Using the more usual notation in which
the coefficient $C_i$ is set to $4\pi$, the associated lower bound on $\Lambda$
is of order $10$ TeV.  This compares well with the current lower bounds of
order 1-2 TeV derived from FNAL data for both the 4-quark operator
$(\bar\psi_L \gamma^\mu \psi_L)^2$ \cite{fnalfq} and the $(DG)^2$ operator
\cite{ehspcone}.

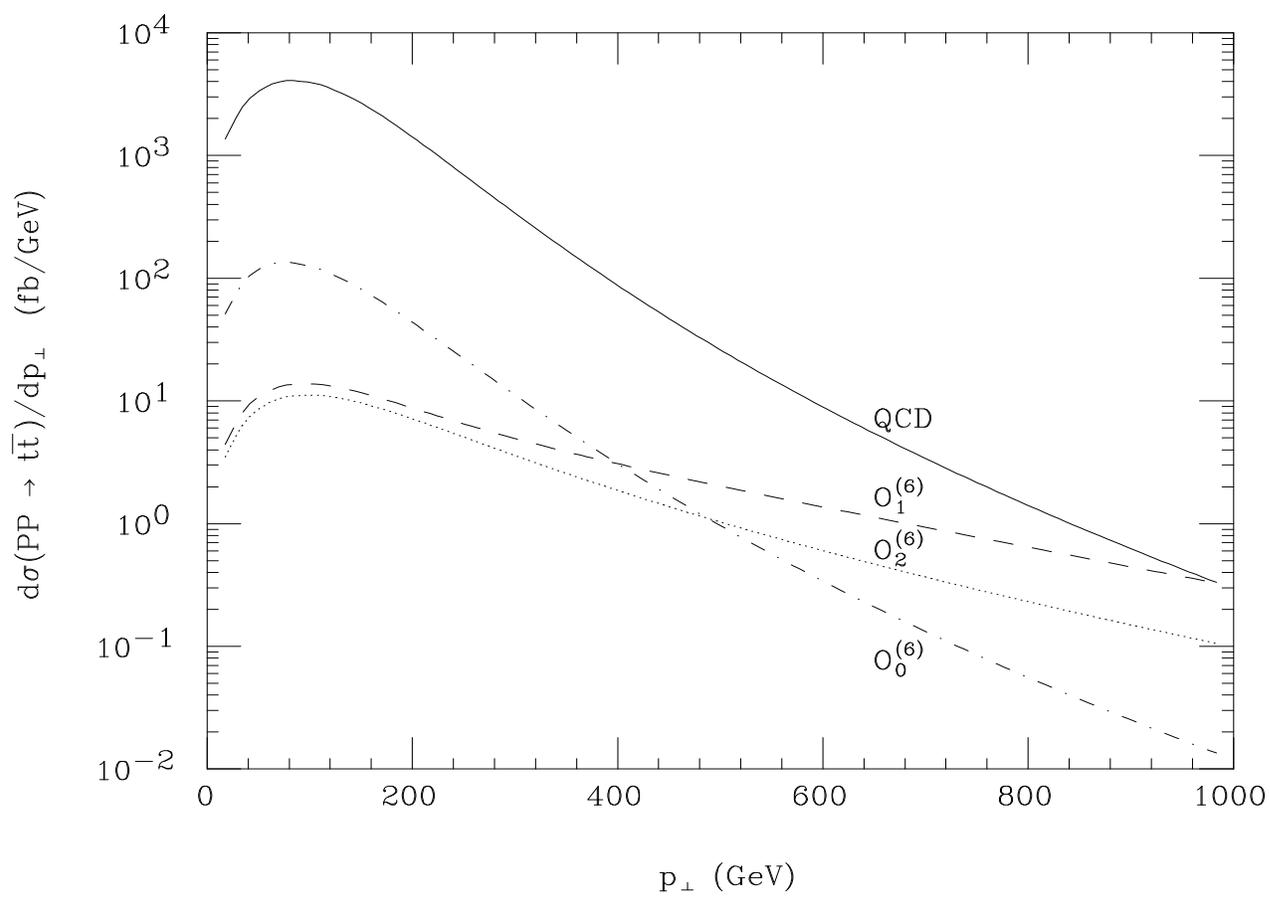
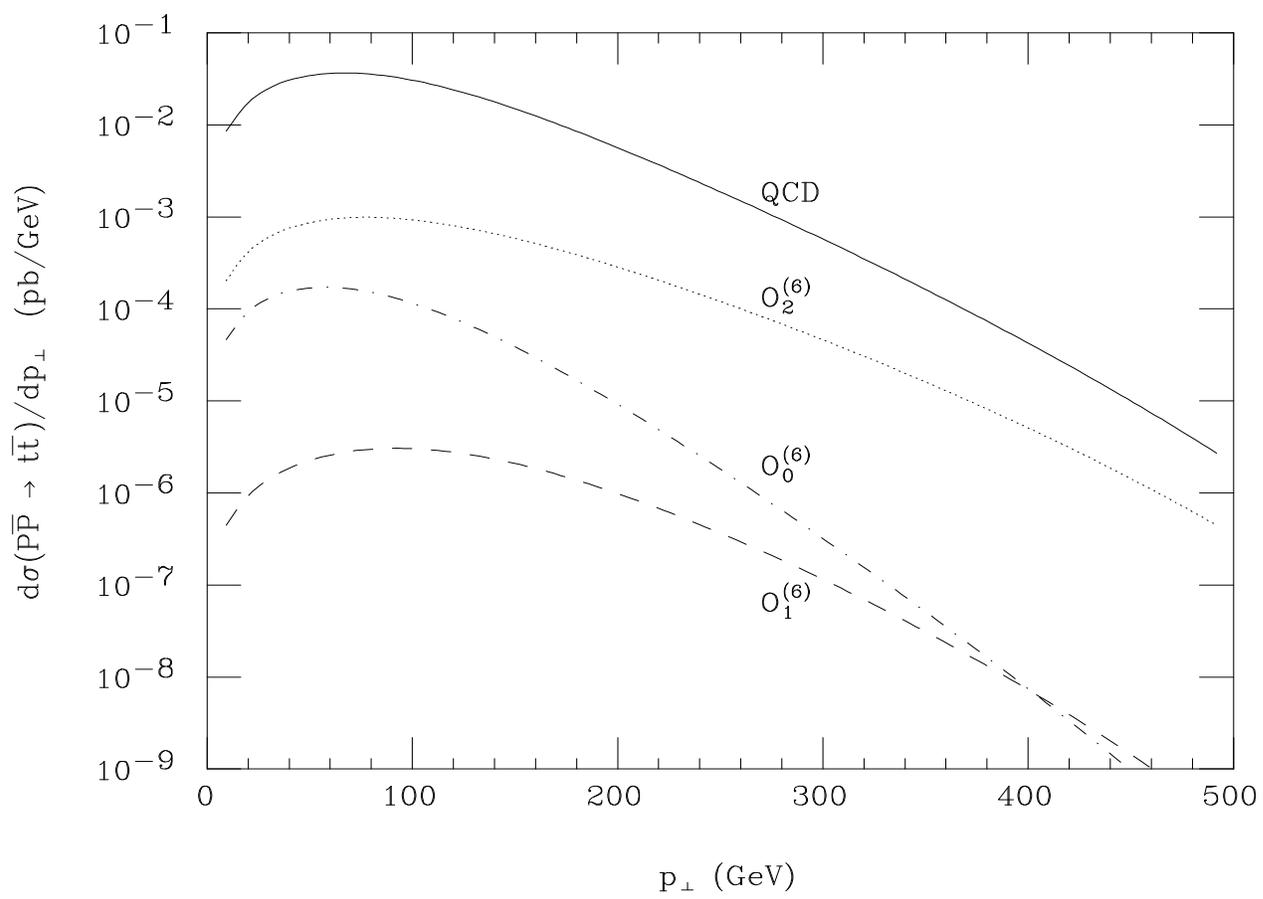
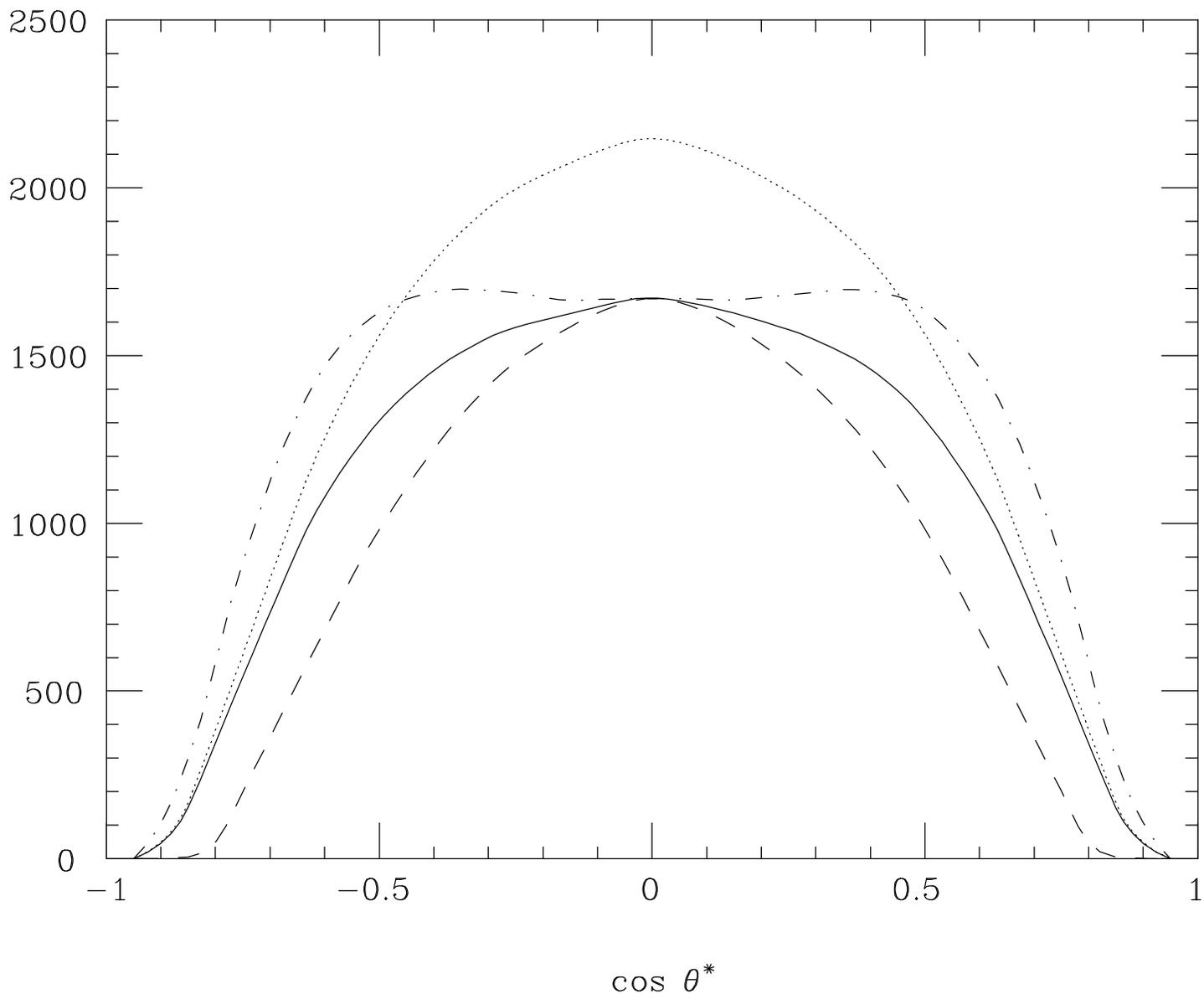
\end{document}